# Science Behind Observations

Slavoljub Mijovic

University of Montenegro; Faculty of Natural Sciences and Mathematics; Cetinjski put 2

20000 Podgorica MONTENEGRO e-mail: slavom@ucg.ac.me

MARK TWAIN

"What gets us into trouble is not what we don't know. It's what we know for sure that just ain't so."

**Abstract:** Global climate change is one of major concern of modern society. To estimate this change usually one estimates the global mean temperature. Measuring and calculating the Earth's average temperature are multi-steps complex processes which combine data from various sources and use statistical techniques. Nowadays, the dataset containing the spatial-temporal data about Earth's temperature is readily to use. Although scientists claim that to be able to achieve an accuracy of a few tenths of a degree, the main question is not the accuracy but, does the global mean temperature make sense at all? It is shown that existing methodology of determining the global mean temperature is quite inappropriate for the estimation of climate change and in long term creating wrong science. A new methodology is introduced, concerning energy budget of heating and cooling of the Earth. The total influence of the atmosphere in global warming can be easily estimate by comparison with the Moon's temperature as a bare body. 'The potential temperature for cooling' is introduced as a right parameter to estimate global warming trend and climate change. Using the new methodology in temperature's averaging, many surprising results would be expected and would give right insight about the climate change.

## Introduction

By climate system we primarily mean the land, ocean, ice on the surface of the Earth together with the atmosphere that overlies it and the radiation from the Sun that provides energy. All of these interact, to produce the conditions on and around the surface of our planet that we call the climate [1]. Averaging physical quantities which characterize these conditions in space and time defines the climate. Typically these statistics are calculated over a period of 30 years [2]. The physical quantities taken into account are the surface temperature but also precipitation, cloud cover, wind field, etc.

The major task as Klaus Hasselmann, Nobel Laureate for physics in 2021, said is the detection problem, often viewed as a task of identifying the most sensitive climate index, from a large set of potentially available indices, for which the anticipated climate change signal can be most readily distinguished from the natural climate noise. Global or regional mean surface temperature, vertical temperature differences, sea ice extent, sea level change, and integrated deep ocean temperatures are examples of indices [3]. Anyway, measuring and reconstructing the global mean temperatures remain the main task of many world's prominent organizations, like NASA *Goddard Institute for Space Science, National Oceanic and Atmospheric Administration* (NOAA) and *the Hadley Centre of the UK Met-Office* [4]. These organizations analyze the data to determine the average temperature of the Earth's surface over a specific period, usually on an annual basis or over longer time spans.

Measuring and calculating the global mean temperature involves collecting temperature data from various locations around the world and then averaging these values to get the overall temperature. Basically, the globe is divided in many space cells and for each grid cell the anomaly- the difference of the measured and the usual temperature on that day, is calculated. Finally, the average of all anomalies is calculated and compared with other years. It's important to note that this process can be complex and requires sophisticated methods to handle issues like data quality, spatial and temporal gaps, and biases. Further, it's worth mentioning that the calculation of the global mean temperature is an ongoing scientific endeavor, and various organizations might use slightly different methodologies and datasets, leading to some variations in the reported global mean temperature values.

Nowadays, the knowledge about the composition and structure of the atmosphere and ocean and their coupled behavior, together with new global measurement systems like remote sensing, and numerical computer models, make possible quantitative studies and predictions of issues such as global warming and climate change [1-6]. For example, only IPCC Sixth Report-Climate Change 2021 was based on an assessment of over 14,000 scientific publications and in IPCC Fifth Report stated, with high confidence, that human-induced warming reached approximately 1°C (likely between 0.8°C and 1.2°C) above pre-industrial levels in 2017, increasing at 0.2°C (likely between 0.1°C and 0.3°C) per decade [1]. Such results, based on the models and the observations, are impressive but at the same time questionable. Although scientists claim that to be able to achieve an accuracy of a few tenths of a degree [1,5] the general public concern and scientific debate still continue.

The focus in this paper is to analyze and answer a simple question: Does the global mean temperature make sense at all? Here is shown that local measured temperatures on the Earth's surface are misused, by the current way, is inadequate and can be used for completely different purposes, but not as a key parameter in the assessment of climate change. At the very least, it is necessary to determine some more temperatures and then compare them.

**THE CURRENT METHOD**

By summing and thereafter averaging the temperatures in different regions of the Earth's surface, we get a quantity which has no direct physical meaning. To say in another way, the given quantity is a pure statistical indicator insensitive to climate change and so far is interpreted in a wrong way. Below are listed examples (thoughtful experiments) that support this claim. The averaging is wrong.

## Thoughtful Experiments

### A Simple Proof about Inadequacy of the Current Methodology

Imagine that the Earth is heated in a such way that has uniformly distributed moderate temperature, $t_{Earth} = 15°C$. Now, if the Earth experiences severe climate change so, that one hemisphere has the mean temperature, $t_1 = 0°C$, and the other, $t_2 = 30°C$, we still have the same mean temperatures, $t_{Earth} = (t_1 + t_2)/2 = 15°C$. We can conclude that there are infinity number of temperature distributions, which actually reflects climate change, although we calculate the same mean temperature. Insensitivity the mean global temperature to climate change is obvious.

### A Spatial Temperature Distribution is Important

To calculate the Earth's temperature one can use different models but all models are based on energy balance budget between heating the Earth with solar shortwave radiation and the Earth's cooling with long-wave infra-red (IR) radiation. In the simplest way for the Earth without atmosphere and in equilibrium of incoming and outgoing energy fluxes, one can easily calculate the mean temperature of Earth surface as:

$$T_{Earth} = \sqrt[4]{\frac{(1-\alpha)S}{4\sigma}} \approx 255K = -18°C, \tag{1}$$

where are $S = 1.361 kW/m^2$ —solar constant, $\sigma = 5.67 \cdot 10^{-8} Wm^{-2}K^{-4}$ — Stefan-Boltzmann constant, and $\alpha \approx 0.3$ — the Earth's albedo. This temperature, usually named as effective or sometimes radiometric, is of course far away from the real Earth's temperature (13.9°C), emphasizing the role of the atmosphere [6].

Two problems arise immediately here that make these comparisons pointless. First, the albedo is different for a bare planet and planet with atmosphere. Second, such calculated temperature as in formula (1), implies uniformly heated planet that actually never happened in reality.

To underline the importance of spatial temperature distribution, let's consider one such drastic case. Imagine, for the sake of simplicity, that only one side of the planet is uniformly heated and the other side remains cold $T_2 \approx 0K$. The temperature of the first side is easily calculated by the equation (1), only four in the denominator is replaced by two, $T_1 \approx 303K$. So, the global mean temperature, with the same source of heating, is now, $T_{Earth} = (T_1 + T_2)/2 \approx 152K$!!!

### A Temporal Temperature Distribution is Important

Imagine that the planet heated uniformly half of the time with two Suns, $2S$, and half of the time without heating i.e averaging in time, we have one Sun. If we further suppose that the body has small heating capacity and immediately cooling without the heating source, we get again the same results as in the previous example. Thus, one can conclude that any spatial and temporal temperature distribution lead to decreasing the global mean temperature. It does not mean the mandatory cooling of the planet.

## THE PROPOSED METHOD

The basic idea is in finding a link between energy balance and global temperature. Due to complexity of the atmosphere influence, let us firstly analize a 'bare' planet like the Moon. It is at the same distance from the Sun as the Earth, so solar constant is the same. Its albedo is 0.12, thus the average temperature can be easily calculated using the zeroth order model [6]: $T_{Moon} = 270K$. Since the temperature distribution is very inhomogeneous and the cooling of each point on the surface is proportional to the fourth degree of temperature, it is necessary to have these values at every moment of time and at every point on the surface to calculate the outgoing energy flux. So, theoreticaly one can calculate the outgoing energy flux of cooling by integrating in a certain period of time the flux from all points of the surface. At the balance point the total incoming energy flux in a certain period of time is equal to the outgoing energy. The same method is analogously applicable to the Earth.

Fortunately, for the Earth we have the dataset of the surface temperatures with some spatial and temporal steps, readily to use. Because of presence of the atmosphere, the cooling is mainly by convection, but one can introduce 'the potential cooling' by radiation i.e simulate the case of absence the atmosphere. Thus, the formula of the energy equilibrium is:

$$\sum_{i=1}^{N} \Delta S_i \left( \sum_{j=1}^{M} \sigma T_j^4 \Delta t_j \right) = F_{in} S_{Earth} \tau, \qquad (2)$$

Where are $\Delta S_i$ — the area of $i$th cell, $N$ —total number of the grid cells, $\Delta t_j$ — $j$ time step, $M$ —total number of time steps in the chosen time period, $\tau = \sum_{j=1}^{M} \Delta t_j$, $F_{in}[W/m^2]$ —incoming energy flux to the Earth's surface, $S_{Earth} = \sum_{i=1}^{N} \Delta S_i$ —the area of the Earth's surface.

Now, the total influence of the atmosphere, like greenhouse effect, can be easily calculated as the difference between 'the potential cooling energy', given by the term on the left side of the equation (2), and total incoming energy that the Earth would receive in that period of time without atmosphere. In this case the value for albedo is changed to the value of albedo of the Earth surface, $\alpha_{Earth\ surface} \approx 0.2$ [7]. For the constant 'potential cooling energy', the global mean temperature, calculated so far, can serve as a measure for redistribution temperature field around the globe.

So, the proper way to calculate the Earth's temperature in a certain period of time, i.e, to estimate climate change (warming or cooling), is

$$\sum_{i=1}^{N} \Delta S_i \left( \sum_{j=1}^{M} \sigma T_j^4 \Delta t_j \right) = S_{Earth}\ \sigma T_{P.C}^4 \tau. \qquad (3)$$

This temperature, $T_{P.C}$, one can call 'the effective temperature for potential cooling'. Previous parameter-the global mean temperature could serve as an indicator for spatial and temporal temperature redistribution.

## CONCLUSION

Here comes the conclusion that current methodology for estimating global warming and climate change by calculating the global mean temperature is insensitive and not adequate. The main reason is that the mean globe temperature has no direct physical meaning and there is not direct conection with the energy balance. Mathematicaly, this inconsistency is the consequence that the global mean temperature is linear combination of other temperatures and energy budget is higly non-linear and depends on the forth power of temperatures. The proper method is introduced and its testing would give a new insight to climate change with the existing datasets. 'The effective temperature for potential cooling' is quite sensitive to the global warming and the change of concentration of green gases in the atmosphere. The fact that previous dataset can be readily used, enables easy implementation of this method  The reference point could be chosen in the pre-industrial time (1850), i.e. establishing carfully the 'natural' greenhouse effect and evolution of climate change and global worming could be folowed by the change of the 'effective temperature for potential cooling'.